\documentclass{emulateapj}
\usepackage{apjfonts}
\usepackage{graphicx}
\usepackage{amsmath}
\usepackage{natbib}
\usepackage{epsfig}

\newcommand{\jcap}{J.~Cosmol.~Astropart.~Phys.~}
\newcommand{\mn}{{Mon.\@ Not.\@ Roy.\@ Ast.\@ Soc.\ }}
\newcommand{\asta}{{Astron.\@ Astrophys.\ }}

\newcommand{\etal}{{\it et al.~}}
\newcommand{\ie}{{i.e.}}
\newcommand{\etc}{{\it etc.}}

\newcommand{\eg}{{e.g.,~}}

\newcommand{\beq}{\begin{equation}}
\newcommand{\eeq}{\end{equation}}
\newcommand{\ber}{\begin{eqnarray}}
\newcommand{\eer}{\end{eqnarray}}
\newcommand{\lleq}{\lower0.9ex\hbox{ $\buildrel < \over \sim$} ~}
\newcommand{\ggeq}{\lower0.9ex\hbox{ $\buildrel > \over \sim$} ~}
\newcommand{\lsim}{\ \lower-1.5pt\vbox{\hbox{\rlap{$<$}\lower5.3pt\vbox{\hbox{$\sim$}}}}\ }
\newcommand{\gsim}{\ \lower-1.5pt\vbox{\hbox{\rlap{$>$}\lower5.3pt\vbox{\hbox{$\sim$}}}}\ }

\newcommand{\pr}{\prime}

\newcommand{\de}{dark energy}
\newcommand{\De}{Dark energy}

\newcommand{\ld}{\Lambda}

\newcommand{\perts}{perturbations}

\newcommand{\HH}{{\cal H}}
\newcommand{\cs}{c_{s,DE}^2}
\newcommand{\ca}{c_{a,DE}^2}

\newcommand{\ww}{w_{DE}}

\newcommand{\w}{w_0}
\newcommand{\wm}{w_m}
\newcommand{\at}{a_t}
\newcommand{\dt}{\Delta_t}

\newcommand{\omt}{\Omega_{0 \rm m}}

\newcommand{\omk}{\Omega_{\kappa}}
\newcommand{\omch}{\Omega_c h^2}
\newcommand{\ombh}{\Omega_b h^2}
\newcommand{\sig}{\sigma_8}

\begin{document}

\title{Galaxy Clusters as a probe of early dark energy}

\author{Ujjaini Alam}
\affil{ISR-1, ISR Division, Los Alamos National Laboratory, Los Alamos, NM 87545, USA}
\email{ujjaini@lanl.gov}
\author{Zarija Luki\'c}
\affil{T-2, T Division, Los Alamos National Laboratory, Los Alamos, NM 87545, USA}
\email{zarija@lanl.gov}
\author{Suman Bhattacharya}
\affil{T-2, T Division, Los Alamos National Laboratory, Los Alamos, NM 87545, USA}
\email{sumanb@lanl.gov}
\thispagestyle{empty}

\sloppy

\begin{abstract}
We study a class of early dark energy (EDE) models, in which, unlike
in standard dark energy models, a substantial amount of dark energy
exists in the matter-dominated era. We self-consistently include dark
energy perturbations, and show that these models may be successfully
constrained using future observations of galaxy clusters, in
particular the redshift abundance, and the Sunyaev-Zel'dovich (SZ)
power spectrum. We make predictions for EDE models, as well as
$\ld$CDM for incoming X-ray (eROSITA) and microwave (South Pole
Telescope) observations.  We show that galaxy clusters' mass function
and the SZ power spectrum will put strong constraints both on the
equation of state of \de~today and the redshift at which EDE transits
to present-day $\ld$CDM like behavior for these models, thus providing
complementary information to the geometric probes of dark energy.  Not
including perturbations in EDE models leads to those models being
practically indistinguishable from $\ld$CDM. An MCMC analysis of future
galaxy cluster surveys provides constraints for EDE parameters that
are competitive with and complementary to background expansion
observations such as supernovae.
\end{abstract}

\maketitle

\section{Introduction}\label{intro}

Over the last decade, observational evidence has mounted in favor of
\de, the mysterious component which dominates the energy content of
the universe at present, and causes the expansion of the universe to
accelerate \citep{sne, wmap7}.  The nature of \de~is one of the most
tantalizing mysteries of present day cosmology. The simplest model for
\de, the cosmological constant, fits the current data well
\citep{sne}, however, there are no strong constraints on the time
evolution of \de~at present. Thus, evolving models of \de~remain as
alternative candidates for dark energy. Many non-cosmological constant
models have been suggested for \de, including scalar field
quintessence models, modifications of the Einstein framework of
gravity, \etc \citep[see reviews][and references therein]{rev1, rev2,
  rev3, rev4, rev5, rev6, rev7}, but as of now there is no clear
consensus on the nature on \de.  An interesting class of models which
have been suggested in the literature are early dark energy (EDE)
models, in which the early universe contained a substantial amount of
dark energy. Geometric probes of \de~put strong constraints on the
present-day nature of \de, and these constraints are expected to
improve with future surveys. However, very little is known as to the
nature of \de~at early times due to paucity of data beyond redshifts
of few. It has thus been hypothesized that even if \de~at present
behaves like the cosmological constant, in the past it could have had
completely different behavior, leading to the idea of the EDE
models. Different facets of these models have been studied in recent
works \citep{ede1,ede2,ede3} and references therein, and have been
analyzed with respect to observations extensively in recent times in
\citep{ede4,ede5,ede6,ede7,ede8, ede9, ede_cmb, ede10}. Current data
places some constraints on these models, but does not rule them
out. In this work we use a parameterization of the equation of state
of \de~to study the possibility of constraining EDE models using
future large scale structure surveys.

Ground and space-based telescopes targeting clusters in microwave
[Atacama Cosmology Telescope (ACT) \citep{act}, the South Pole
Telescope (SPT) \citep{spt} and Planck
\footnote{http://www.rssd.esa.int/index.php?project=Planck}] have
begun operation, or will return data shortly. Also current (Ebeling et
al.~2010) and future missions have been planned for detecting clusters
in X-ray waveband (Predehl et al.~2007). It has been recognized for
some time now, that galaxy cluster surveys in X-ray or microwave via
Sunyaev-Zel'dovich (SZ) effect could be precision probes of
cosmological parameters, in particular the dark energy density and its
equation of state parameter \citep{haiman2001}. This is because
redshift distribution of massive clusters is exponentially sensitive
to the growth of structure history of the universe, which in turn,
bears the signature of dark energy. The presence of early dark energy
reduce structure formation, consequently the cluster abundances
reduce. Thus cluster counts can provide a smoking gun probe for
detecting early dark energy. For example, \cite{mantz09a} have used a
flux limited X-ray cluster data to constrain early dark energy
component assuming the redshift of transition between 0 to 1. The
current generation of SZ surveys will also measure the power spectrum
of the cosmic microwave background with an unprecedented accuracy down
to scales of 1 arc-min. The SZ surveys have started measuring the SZ
power spectrum which is the dominant signal at scales of few $arc-min$
\citep{sptcl, actcl}. The amplitude of the SZ power spectrum is
proportional to the total number of objects that have formed and hence
is extremely sensitive to the amount of dark energy present at early
epoch.

The paper is organized as follows. In Section~\ref{ede} we explain the
early dark energy formalism and model used for this
analysis. Section~\ref{obs} expounds the nature of X-ray and microwave
observations and mass-observable scaling relations. In
section~\ref{res} we show predictions from current and future
observations, comparing our EDE models to a fiducial $\Lambda$CDM
cosmology. Finally, section~\ref{concl} is devoted to conclusions and
discussion.

\begin{figure*}[t] 
\includegraphics[height=8cm, width= 16cm]{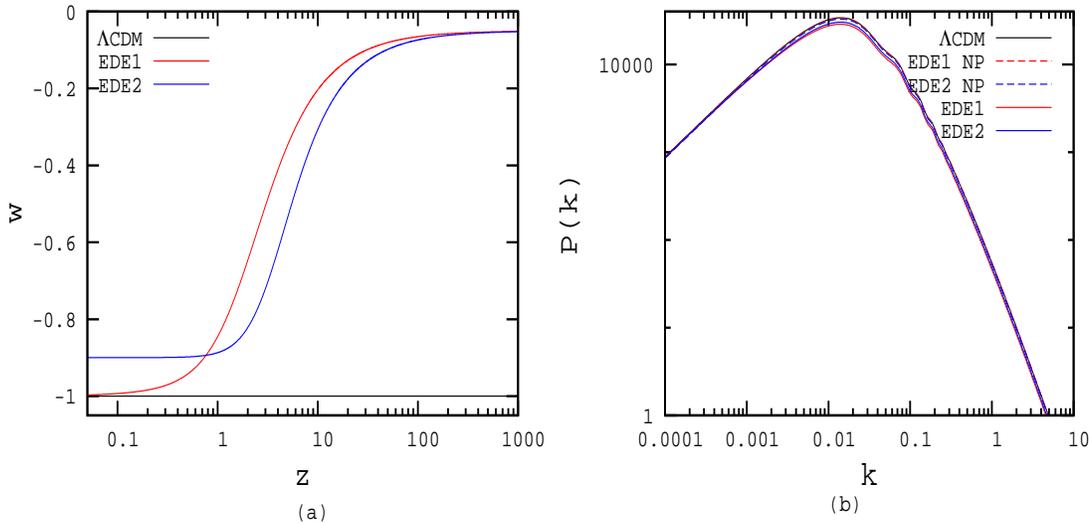}
\caption{\small 
Equation of state (panel (a)) and matter power spectrum at $z=0$
(panel (b)) for EDE models with $\w = -1.0, \wm = -0.05, \at =0.17,
\dt =0.17$ (EDE1, red lines), and with $\w = -0.9, \wm = -0.05, \at
=0.1, \dt -=0.1$ (EDE2, blue lines). The black line in each panel
represents the corresponding $\ld$CDM model, in panel (b), the dashed
lines represent EDE without DE perturbations, while the solid lines
represent EDE with perturbations. }
\label{fig:ede}
\end{figure*}

\section{Early Dark Energy}\label{ede}

\De~\perts~for dynamic \de~models have been studied in a number of
works, usually under the formalism of a minimally coupled scalar field
\citep[See][and references therein]{pert1, pert2, pert3, pert4, pert5,
  pert6, pert7, pert8, pert9, pert10}.  In this work we follow the
formalism of \cite{pert5}. First order perturbations in a homogeneous
and isotropic large scale universe described by the
Friedman-Lemaitre-Robertson-Walker (FLRW) metric take the form
\beq
ds^2 = a^2(\eta) \left[(1+2\Psi({\bf x}, \eta)) d\eta^2 - (1+2\Phi({\bf x}, \eta)) \delta_{\alpha\beta}dx^{\alpha} dx^{\beta} \right]\,\,, 
\eeq 
where $\eta$ is the conformal time, ${\bf x}$ is the length element,
$a(\eta)$ is the scale factor, and $\Phi, \Psi$ are the Bardeen
potentials. If proper isotropy of the medium is zero, then $\Phi =
-\Psi$.

Along with the matter and radiation components, we consider dark
energy to be an additional fluid component, so that the dark energy
\perts~are characterized by an equation of state and an adiabatic sound
speed--
\ber
\ww &=& \frac{p_{DE}}{\rho_{DE}} \\
\ca &=& \frac{\dot{p}_{DE}}{\dot{\rho}_{DE}}  \,\,.
\eer
Defining the frame-invariant quantity $c_{s,i}^2$ (the fluid sound
speed in the frame comoving with the fluid), the evolution equations
for a fluid with equation of state $w_i = p_i/\rho_i$, and adiabatic
speed of sound $c_{a,i}^2 = \dot{p}_i/\dot{\rho}_i$, can be written as
(prime denotes derivative with respect to $\eta$)
\ber\label{eq:pert}
\delta_i^{\pr} &=& -3 \HH (c_{s,i}^2-w_i) \delta_i - 9 \HH^2 (c_{s,i}^2-c_{a,i}^2)  (1+ w_i) \frac{v_i}{k}  \nonumber\\
&& -(1+ w_i) k v_i - 3(1+ w_i) \Psi^{\pr} \\
v_i^{\pr} &=& - \HH (1-3 c_{s,i}^2) v_i + \frac{k c_{s,i}^2 \delta_i }{(1 + w_i)} - kA  \label{eq:pert2}\,\,, 
\eer
where $\HH = a^{\pr}/a$ is the conformal Hubble parameter and $A$ is
the acceleration ($A = 0$ in the synchronous gauge, $A = -\Psi$ in the
Newtonian gauge). Adiabatic initial conditions are considered.  For
the matter component, $w_m = c_a^2 = c_s^2 = 0$. For the dark energy
component, a fluid with varying $\ww \geq -1$ has $\ca = \ww -
[d\ww/d({\rm ln} \ a)]/3(1+\ww)$. For scalar field like \de~models,
$\cs = 1$. For a more general class of models, such as k-essence,
$\cs$ could be variable as well.

If we consider dark energy without perturbations, the quantities
$\delta_{DE}, \delta^{\pr}_{DE} = 0$, and the matter density contrast
is given by
\beq\label{eq:nopert}
\delta_m^{\pr \pr} - \HH \delta_m^{\pr} -4 \pi G \rho_m \delta_m = 0 \,\,,
\eeq
thus the dark energy component appears only in the damping term, so
that a non-negligible amount of dark energy would lead simply to a
suppression of clustering of matter at large scales. Not taking into
account the dark energy perturbations can lead to gauge dependent
results, as shown in \cite{pert_gauge}, in this paper we have used the
commonly used comoving gauge. If dark energy perturbations are
included self-consistently, the results are gauge-independent.

To study EDE models under this formulation, we consider a
$w$-parameterization which may represent a large class of varying
\de~models \cite{coras}
\beq
w(a) = \w +(\wm-\w)\frac{1+e^{\at/\dt}}{1+e^{-(a-\at)/\dt}}\frac{1-e^{-(a-1)/\dt}}{1-e^{1/\dt}} \,\,,
\eeq
where $\w$ is the equation of state of dark energy today, $\wm$ is the
equation of state in the matter dominated era, $\at$ is the scale
factor at which the transition between $\w$ and $\wm$ takes place, and
$\dt$ is the width of the transition. For studying EDE models with
this parameterization, we choose $\wm > -0.1$, to ensure the presence
of adequate amount of \de~at early times.

Current data puts impressive constraints on the values of these
parameters \citep[see][]{ede_cmb}, however, many interesting EDE models
still fall in the acceptable range. We choose two such EDE models in
order to study the possibility of constraining these models further
using future observations. The models chosen are
\begin{itemize}
\item Model 1 : $\w = -1.0, \wm = -0.05, \at = 0.17, \dt = 0.17$;
\item Model 2 : $\w = -0.9, \wm = -0.05, \at = 0.1, \dt = 0.1$.
\end{itemize}
Model 1 behaves like $\ld$CDM at present, while Model 2 has a higher
value of the equation of state at present. We compare the results for
these models with those for a $\ld$CDM model with identical values for
the non-\de~parameters. The non-\de~parameters are chosen from the
WMAP7 CMB+BAO+HST best-fit, \eg $\ombh = 0.0226, \omch= 0.1123, h =
0.704, n_s =0.963$, as well as other parameters such as the scalar
amplitude $A_s$ and the reionization optical depth $\tau$
\citep{wmap7}. We modified the publicly available COSMOMC code
\cite{cosmomc} for generating the transfer functions, and therefore
the power spectrum at different redshifts for these models.

Figure~\ref{fig:ede} shows the behavior of the equation of state of
\de, and the matter power spectrum at redshift $z=0$ for these two
models. For comparison we also plot the results for the matter power
spectrum when \de~perturbations are not considered. We see that when
the perturbations are not considered, the EDE matter power spectrum,
although there is a suppression the matter power spectrum, this effect
is negligible, and for both models the matter power spectrum is very
close to that for $\ld$CDM. However, when the perturbations are
considered, the matter power spectrum is significantly different from
$\ld$CDM, resulting in a $\sig = 0.74$ for EDE1, and a $\sig = 0.76$
for EDE2, whereas the $\ld$CDM has a $\sig =0.81$ . This behavior is
typical of EDE models with rapid and large transition of the equation
of state. This translates to a rapid change in the \de~\perts~at high
scales (low $k$), thus to enhancement in the transfer function at low
$k$. When normalized to the CMB scale, this effect shows up as an
apparent suppression of power at high $k$, and therefore a low
$\sig$. The effect is stronger in EDE1 (\ie~lower $\sig$) because the
difference between the equation of state at present and in the matter
dominated era in this case is larger. Further details on how the
change in equation of state affects the dark energy perturbations can
be found in paper I \cite{ede_cmb}.

\section{Observations}\label{obs}

Paper I \cite{ede_cmb} explores effects of early dark energy on CMB power
spectrum and other current observations; here we focus on forthcoming
large-scale structure probes. We consider the influence of EDE on cluster
counts and SZ power spectrum.

\subsection{Cluster Counts}
\label{counts}

Redshift evolution of cluster abundance provides a valuable insight
into global dynamics of the universe. This abundance is sensitive to
both expansion and growth history of the universe, and as such can be
used to constrain standard cosmological parameters like $\sigma_8$ or
$\omt$ (see, e.g. Haiman et al.~2001), but also to explore possible
mechanisms for observed acceleration, like modifying the law of
gravity.

For a spatially flat cosmologies considered here ($\omk \equiv 0$),
all-sky number of objects more massive than $M_{min}$ in a redshift
bin is:
\begin{equation}
N(z) = \int_{M_{min}}^{\infty}  \int_{z_1}^{z_2} \frac{4 \pi r^2(z) dr(z)}{dz} \frac{dn(M,z)}{dM} dz \, dM \ ,
\end{equation}
where the first term is the comoving volume, and the second is the
differential mass function.

\begin{figure}[t]
\includegraphics[width=8cm]{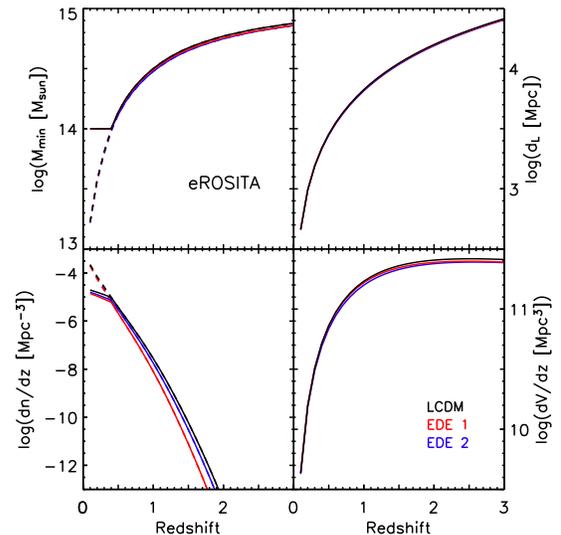}
\caption{
eROSITA survey, from upper left panel in clockwise direction: minimal
detectable mass (dashed line shows instrument capability, solid line
shows minimal mass for galaxy clusters only with mass threshold of
$10^{14} M_{\odot}$), luminosity distance, comoving volume element
covered by the survey, and comoving cluster abundance as a function of
redshift. 
}
\label{fig:erosita}
\end{figure}

From numerical simulations, we have a good quantification of the halo
mass function. Although its precision is still not at the level needed
to make it insignificant for future cosmological parameter studies
(Cunha \& Evrard 2009; Wu et al.~2009), that will not be relevant here
as we want to explore the relative difference between our EDE models and
the fiducial $\Lambda$CDM cosmology. The important point, on which we rely
here, is that the mass function can be presented in a {\it universal}
form, where the cosmology dependence comes from the linear power
spectrum, and the linear growth factor. At a 20\% level (often better
than that), it is indeed proven to be the case, for many different
cosmological models (Jenkins et al.~2001; Linder \& Jenkins 2003;
Jennings et al.~2010; Bhattacharya et al.~2010) and for a wide range
of redshifts (Heitmann et al.~2006; Luki\'c et al.~2007; Reed et
al.~2007). Since we are interested in modeling X-ray and SZ
observations, we will use mass function of halos defined as spheres
enclosing a given overdensity, defined with respect to the critical
density for the closure, $\rho_c$ (for analysis of different halo
definitions see, e.g. White 2002; Luki\'c et al.~2009).

One way of detecting groups and clusters of galaxies is via weak
lensing maps (Marian \& Bernstein 2006); this approach is appealing as
masses are probed directly there, but on a downside the method can
probe only the most massive systems, and only within limited redshifts
(as galaxy shapes are very hard to measure beyond $z \sim 1$).
Promising alternatives to the weak lensing are mass measures through
Sunyaev-Zel'dovich effect (e.g. Carlstrom et al.~2002), X-ray emission
(e.g. Kravtsov et al.~2006), and galaxy richness (High et
al.~2010). For these, indirect mass measurements, one first has to
connect cluster mass to a relevant observable for a given survey, like
X-ray flux \citep{mantz09} or integrated Compton y-parameter
\citep{V10, sehgal10}.  In this paper we will focus on X-ray and SZ
surveys, and in the following we explain scaling relations we use.

In the self-similar theory (Kaiser 1986, 1991), temperature of the
intracluster medium scales with gravitational potential ($T \propto
M/R$), and enclosed mass is $M \propto R^3$. This self-similar scaling
was confirmed in hydrodynamical simulations (Mathiesen \& Evrard 2001;
Borgani et al.~2004; Kravtsov et al.~2006), as well as in observations
(Vikhlinin et al.~2009). In the following, we will use $M-T$ relation
from Mathiesen \& Evrard (2001):
\begin{equation}
\frac{M_{200}}{M_{\odot}h{-1}} = 10^{15} \left( \frac{kT}{4.88 {\rm keV}} \right)^{3/2} E(z)^{-1}\ ; \ \ \ E(z) \equiv H(z)/H_0 \ .
\label{MT_relation}
\end{equation}
We use the above relation, as it relates $M_{200}$ to the bolometric
luminosity. While X-ray scaling relations are much tighter for
$M_{500}$ (Kravtsov et al.~2006), here we want to use the same mass
for both X-ray and SZ estimates, and whether it is the best
observational approach is not of concern here.

Self-similar theory fails to correctly predict $T-L$ relation
(Markevitch 1998; Allen \& Fabian 1998), and thus mass-luminosity
relation is inaccurate as well. The departure from self-similarity is
due to excess entropy in cluster cores which prevents gas from being
compressed to very high densities (Ponman et al.~1999, Finoguenov et
al.~2002). Therefore, in the following we will use $M-L$ relation from
Bartelmann \& White (2003):
\begin{equation}
L = 1.097 \times 10^{45} \left[ \frac{M_{200}}{10^{15}M_{\odot}h^{-1}} E(z) \right]^{1.554} {\rm erg/s}
\label{LM_relation}
\end{equation}

\begin{figure}[t]
\includegraphics[width=8cm]{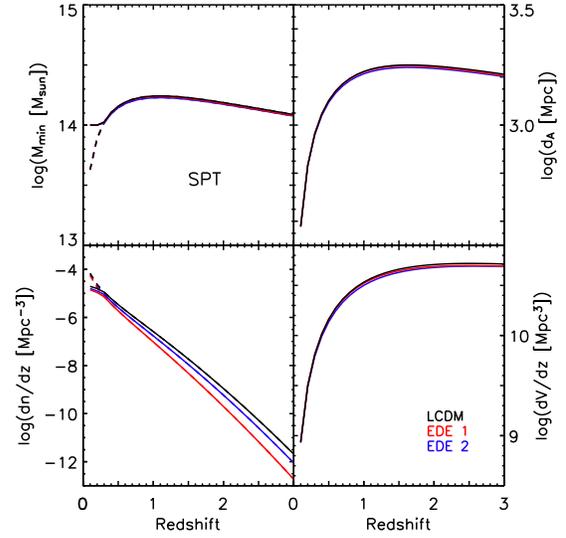} 
\caption{
South Pole SZ survey, from upper left panel in clockwise direction:
minimal detectable mass (dashed line shows instrument capability --
same as in Figure \ref{fig:erosita}), angular diameter distance,
comoving volume element covered by the survey, and comoving cluster
abundance as a function of redshift.}
\label{fig:spt}
\end{figure}

X-ray observations are sensitive in a given energy band, whose flux is: 
\begin{equation}
S_b = \frac{L_X f_b}{4\pi d_L(z)^2} \ ,
\end{equation}
where $f_b$ is the band correction and $d_L(z)$ is luminosity distance
to a cluster:
\begin{equation}
d_L(z) = \frac{c(1+z)}{H_0} \int_0^z \frac{dz'}{E(z')} \ .
\end{equation}
Emission from the intracluster medium is predominantly thermal
bremsstrahlung, but for $T < 2{\rm keV}$ line emission from metals
(clusters have metallicity $\sim 0.3 Z_{\odot}$) becomes
non-negligible (Sarazin 1986). Thus, to calculate $f_b$ we use XSPEC
package (Arnaud 1996) and we assume Raymond-Smith (1977) plasma model
for cluster emission. As the band correction depends on the plasma
temperature, we solve iteratively for minimum mass and corresponding
temperature using XSPEC and equation \ref{MT_relation}. Galactic
absorption is modeled with constant column density of $n_H = 10^{21}
{\rm cm^{-2}}$, roughly corresponding to Galactic mid-latitudes.  In
observational analysis one would use appropriate $n_H$ for the line of
sight to each cluster, and column density will generally be smaller
than our number for high latitudes, and larger for low latitudes.

For the Sunyaev-Zel'dovich observations, one has to relate Compton y
parameter (integrated over the solid angle) to the mass of an
object. Furthermore, it is necessary to renormalize such relation to
the directly observable SZ flux. In Fedeli et al.~(2009), this was
done using Sehgal et al.~(2007) $Y_{200}-M_{200}$ relation, and
assumption that SZ signal outside cluster virial radius can safely be
neglected, resulting in relation:
\begin{equation}
S_{\nu} = \frac{6.763 \times 10^7 {\rm mJy}}{(d_A(z)/1 Mpc)^2} \left( \frac{M_{200}}{10^{15}M_{\odot}} \right)^{1.876} | f(\nu) | E^{2/3}(z) \ .
\end{equation}
Here, $d_A$ is angular diameter distance to the object, and $f(\nu)$
is spectral signature of the thermal SZ effect:
\begin{equation}
f(\nu) = \frac{x^4 e^x}{(e^x-1)^2} \left[ x \frac{e^x+1}{e^x-1} - 4 \right]; \ \ \ x \equiv \frac{h\nu}{kT_{\gamma}}\ ,
\end{equation}
and $T_{\gamma}$ is the CMB temperature. 

\subsection{Sunyaev-Zeldovich Power Spectrum}
\label{szcl}

The current generation of CMB experiments will measure the power
spectrum of the cosmic microwave background with an unprecedented
accuracy down to scales of 1 arc-min. While the primary CMB
fluctuations dominate the power spectrum at a degree scale, at scales
of few arcminutes the secondary fluctuations arising from the SZ
effect and lensing of the CMB becomes the dominant signal.

Predictions for the SZ power spectrum amplitude $C_{l,SZ}$ (henceforth
$C_l$) can be made using the halo model and estimates of the radial
pressure profile of intra-cluster gas. Assuming that the cluster gas
resides in hydrostatic equilibrium in the potential well of the host
dark matter halo, \cite{ks02} demonstrated that the ensemble averaged
power spectrum amplitude $C_l$ has an extremely sensitive dependence
on $\sigma_8$ where $C_l \propto \sigma_8^7 (\ombh)^2$. The kinetic SZ
power spectrum contribute roughly 10\% to the total SZ angular power
spectrum and hence we consider only the thermal component of the SZ
power spectrum (tSZ hereafter) here. The cosmological information of
the tSZ power spectrum comes from the total number of halos of mass $M
\geq 10^{13} M_{\odot}/h$ in the survey area. Thus the SZ power
spectrum contains a significant contribution from low mass clusters
and group-mass objects. Note that measurements of the mass function
usually probe only massive halos ($M \geq 10^{14} M_{\odot}$) and not
group size halos.

The SZ power spectrum can be calculated by simply summing up the
squared Fourier-space SZ profiles, $\tilde{y}(M,z,\ell)$ of all
clusters:
\begin{equation}
C_{\ell} =  g_\nu^2 \int dz {dV \over dz } \int d \ln M {dn(M,z) \over d \ln M}
\tilde{y}(M,z,\ell)^2
\end{equation}
where V(z) is the comoving volume per steradian and $n(M,z)$ is the
number density of objects of mass $M$ at redshift $z$. $g_\nu$ is the
frequency factor of the SZ effect. In this study, we show the power
spectrum at 150 GHz where $g_\nu=-1$. Note that whilst this
calculation assumes that clusters are Poisson distributed,
\citep{shaw09} have shown that including halo clustering modifies the
power spectrum (compared to the Poisson case) by less than 1\%.

The number density of halos $n(M,z)$ can be calculated for a $\Lambda
CDM$ cosmology using the fitting functions provided by \citep{J01} and
more recently \citep{tinker08}.  For the SZ profiles (also mass and
redshift dependent), previous studies have frequently used either a
simple $\beta$-model or the hydrostatic model of \cite{ks02}. Note
that both models simply assume that the gas resides in hydrostatic
equilibrium in the potential well of an NFW-like halo (and is
isothermal in the case of the $\beta$-model. Neither model accounts
for the fraction of hot gas that will have cooled and been converted
into stars or any non-thermal energy input into the ICM. As shown in
\cite{mwhite01, sb07, cita10}, these effects change the SZ power
spectrum by 10-30\%. One possible way to include the uncertainty in
gas physics is the semi-analytic approach of modeling gas physics
\citep{bode07, bode09, shaw10} and calibration the parameters using
X-ray observations of individual clusters. The other source of
uncertainty is the non-gaussian nature of the SZ power spectrum
\citep{shaw09}. All these effects needs to be accounted for precision
modeling of the power spectrum.  This will be essential to harness the
full merit of the upcoming datasets. In the current study we choose to
use the model by \cite{ks02} to assess the impact of dark energy on
the SZ power spectrum.

\section{Results}\label{res}

\subsection{Cluster counts}
\label{res:counts}

\begin{figure}[t]
\includegraphics[width=8cm]{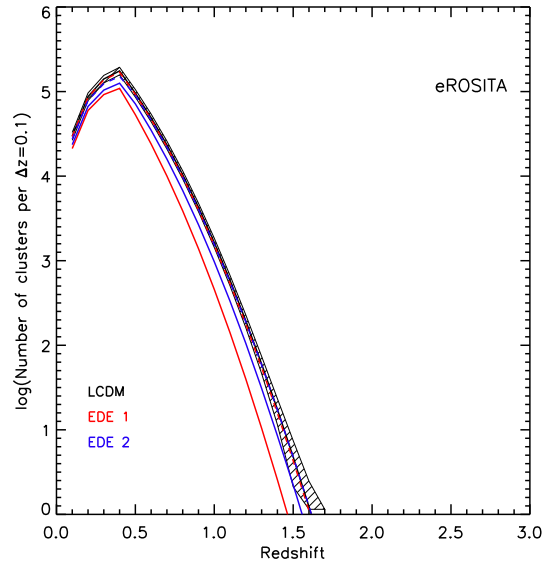} 
\caption{
Expected redshift distribution of galaxy clusters for eROSITA
Telescope. In black is our fiducial $\Lambda$CDM model, red and blue
are EDE1, and EDE2, respectively, and in dashed lines we show results
of EDE models when perturbations are neglected. Shaded regions are
statistical Poisson errors plus 10\% systematic error coming from
non-universality of the mass function.
}
\label{fig:erosita_mf}
\end{figure}

We consider two observational campaigns: upcoming X-ray survey
eROSITA, and ongoing South Pole SZ survey. The first one, due in 2012,
will cover almost half a sky ($f_{sky} \approx 0.49$), and have flux
limit of $F_{min} = 3.3 \times 10^{-14} {\rm erg s^{-1} cm^{-2}}$ in
energy band [0.5, 5.0] keV. South Pole Telescope will scan 4000 square
degrees ($f_{sky} \approx 0.1$) with limit of $S_{min} \approx 5 {\rm
  mJy}$ at frequency $\nu_0 = 150 {\rm Hz}$. Using mass-observable
scaling relations described in section \ref{counts} we can find
minimum detectable mass for each survey, as well as redshift volume
element being observed (Figs. \ref{fig:erosita} and \ref{fig:spt}).
For the theoretical prediction of cluster abundance, we use Tinker et
al.~(2008) mass function for the overdensity of $200 \rho_c$ ($\Delta
\approx 778$ in their notation), and we use their fitting formulas for
the parameters of $f(\sigma)$ as the function of $\log \Delta$. Total
number density of objects with masses above the given threshold as
predicted by Tinker et al.~(2008) is shown in lower left panels of
Figures \ref{fig:erosita} and \ref{fig:spt}, for eROSITA and SPT
survey, respectively.

\begin{figure}[t]
\includegraphics[width=8cm]{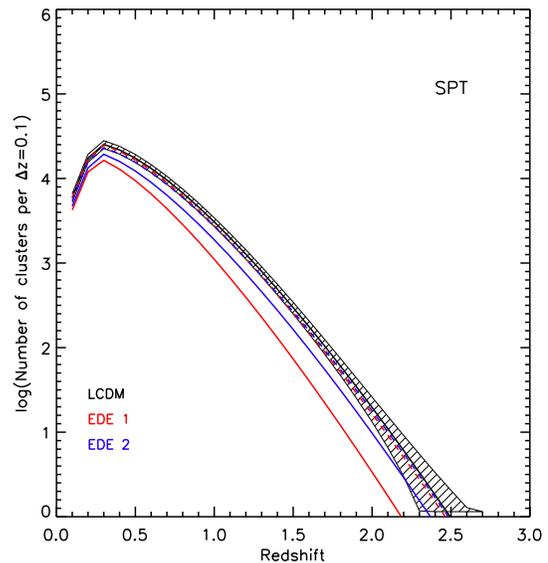} 
\caption{
The same as Fig.~\ref{fig:erosita_mf}, but for the South Pole
Telescope.
}
\label{fig:spt_mf}
\end{figure}

Figures \ref{fig:erosita_mf} and \ref{fig:spt_mf} show expected
redshift distribution of galaxy clusters for the cosmologies
considered here, where we have taken redshift bins of $\Delta z =
0.1$. As expected, SZ survey can detect meaningful number of clusters
up to higher redshift, but has fewer objects at lower redshifts, due
to the smaller sky coverage. Shaded area shows Poisson errorbar plus a
systematic 10\% uncertainty which describes our confidence in
universal formula of the mass function. Dashed lines show expectations
for EDE models if perturbations would be neglected. We see that
presence of perturbations significantly affect growth of structure in
the Universe, effectively reducing $\sigma_8$. Since amount of this
reduction is redshift dependent, it is not possible to mimic it with
$\Lambda$CDM cosmology with different $\sigma_8$ (i.e. it is possible
to mimic it at a particular redshift, but not at a redshift
range). Most importantly, we see that perturbations cannot be
neglected when considering predictions from dynamical EDE models. 

Finally, Figures \ref{fig:erosita_mf} and \ref{fig:spt_mf} showcase
that future large-scale structure results will be able to constrain
EDE models significantly better than is currently the case. Both EDE1
and EDE2 models, which represent models which are (a)
indistinguishable from $\Lambda$CDM at present (EDE1), or (b) at the
upper limit of $w_0$ (EDE2), are distinguishable from $\Lambda$CDM
cosmology with high significance (2$\sigma$ or more). Constraining
power is larger on the EDE1 model which behaves the same as
$\Lambda$CDM today but has a later transition from EDE like behavior
(at redshift $z \simeq 5$), as compared to the EDE2 model which is
different from $\Lambda$CDM today but has a much earlier transition
from EDE like behavior (at redshift $z = 9$). Thus the cluster counts
may constrain not only the current equation of state of dark energy,
but also the redshift at which transition from EDE behavior occurs.

\subsection{SZ power spectrum}
\label{res:sz}

\begin{figure}[t]
\includegraphics[width=8cm]{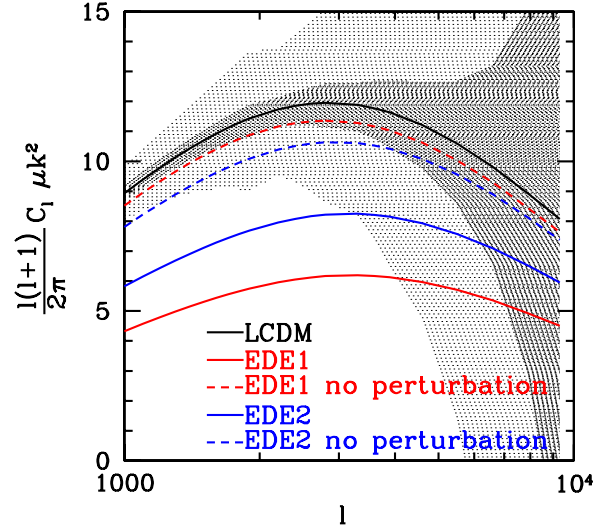} 
\caption{
The impact of dark energy perturbations on the SZ power
spectrum. Different early dark energy models are explained in
section~\ref{ede}. The error bars represent a Planck like full sky
survey and ACT/SPT like higher resolution survey. The error bars
include cosmic variance including the non-gaussian contribution from
the SZ trispectrum and the resolution of the surveys. We assume
perfect removal of astrophysical contaminations from infrared and
radio galaxies and primary CMB.
}
\label{fig:Cl_ede}
\end{figure}

In this section, we show the impact of the dark energy perturbation on
the SZ power spectrum. The presence of perturbations in the dark
energy changes the transfer function, volume factor and the growth
function of the universe, consequently the abundance of the halos
changes. Since SZ power spectrum is proportional to the total
abundance of the halos that have formed in the universe and volume of
the universe, it depends strongly on the perturbations of dark
energy. The impact of different early dark energy models on the SZ
power spectrum is shown in figure~\ref{fig:Cl_ede}. To assess the
merit of the upcoming datasets to constrain early dark energy we
assume two fiducial surveys:\\
1) Planck like full sky survey: We assume 75\% of the full sky
coverage to exclude contaminations due to galactic foregrounds, 7'
resolution and 7 $\mu K$ per pixel as the sensitivity.\\ 
2) ACT/SPT like higher resolution surveys: We assume a coverage of
4000 $deg^2$, 1.2' angular resolution and a sensitivity of 20 $\mu K$
per pixel.\\

The error on the SZ power spectrum for closely separated bins of size
$\Delta l$ in $l$ space can be written as
\ber
\sigma^2(C_l) &=& f_{sky}^{-1} \left[ \frac{2 C_l^2}{(2l+1)\Delta l} \right. \nonumber\\
&&\left. + \frac{2}{(2l+1)\Delta l}\left(f_{sky}w^{-1}\exp\left[\frac{l^2\theta_{fwhm}^2}{8\ln 2}\right]\right)^2  + \frac{T_{ll}}{4\pi} \right] 
\eer
where $f_{sky}$ is the fraction of the sky covered by the assumed
surveys, $w^{-1}=[\sigma_{pix}\theta_{fwhm}/T_{CMb}]^2$,
$\sigma_{pix}$ is the noise per pixel, $\theta_{fwhm}$ is the
resolution at full width at half maximum assuming a gaussian beam,
$T_{CMB}$ is the CMB temperature \citep{jungman96}.

The trispectrum $T_{ll}$ (Poisson term) is given by
\begin{equation}
C_{\ell} =  g_\nu^4\int dz {dV \over dz } \int d \ln M {dn(M,z) \over d \ln M}\tilde{y}(M,z,\ell)^4
\end{equation}

As shown in fig.~\ref{fig:Cl_ede}, the upcoming measurements of SZ
power spectrum will be able to distinguish between $\Lambda$CDM and
different early dark energy models (that are not ruled out by current
observations) with high significance. As in the case of the galaxy
cluster counts, the SZ power spectrum is also sensitive to the
redshift of transition from EDE-like to present day $\ld$CDM-like
behavior. Current constraints on the redshift and width of transition
are $z_t \ggeq 4, \Delta_t \lleq 0.2$ \citep{ede_cmb}, while future
observations of SZ power spectrum may even rule out EDE models with
$z_t \ggeq 9, \Delta_t \lleq 0.1$ (EDE2). These results are of course
for a fixed value of other non-dark energy parameters, and we expect
degeneracies with these in a full analysis. In the next section we
attempt to constrain the EDE parameters using future simulated SZ
power spectrum measurements and cluster counts.

\begin{figure*}[t]
\includegraphics[width=16cm]{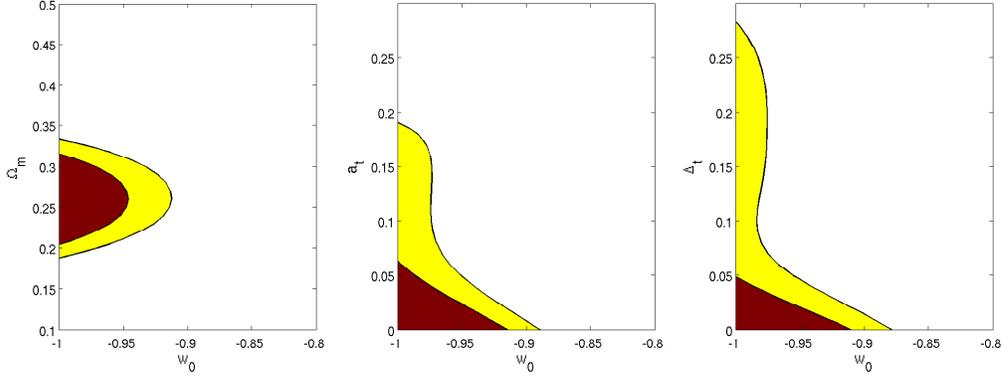}
\caption{Confidence levels for a combination of primary CMB power spectrum from Planck survey + cluster counts obtainable from ACT/SPT survey.}
\label{fig:SPT_mcmc}
\end{figure*}

\begin{figure*}[t]
\includegraphics[width=16cm]{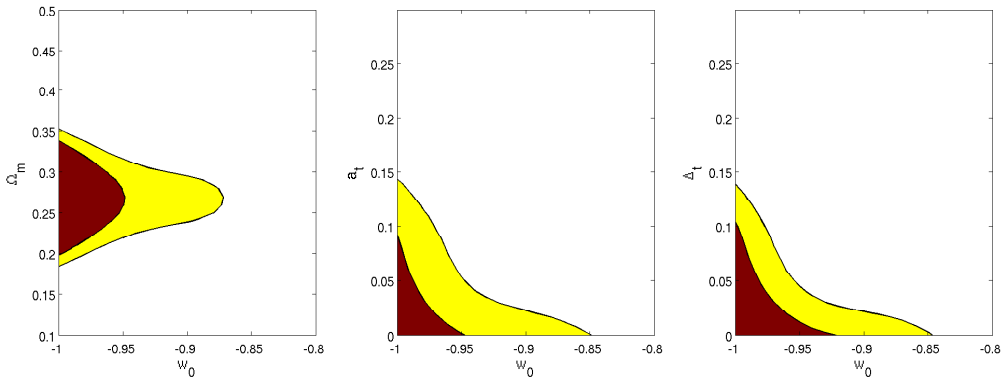}
\caption{Confidence levels for a combination of primary CMB power spectrum from Planck survey +SZ power spectrum obtainable from ACT/SPT survey.}
\label{fig:SZ_mcmc}
\end{figure*}

\subsection{Likelihood analysis}
\label{res:mcmc}

In this section we study the effect of the degeneracies between the
different cosmological parameters for EDE models. We simulate the data
according to WMAP7+BAO+$H_0$ $\Lambda$CDM model \cite{wmap7}. We
generate the primary CMB TT power spectrum simulated to match the
Planck survey resolution and sky coverage. We also simulate the SZ
power spectrum data as obtainable from ACT/SPT surveys with the survey
parameters discussed in detail in section~\ref{res:sz}.  For studying
the cluster data, we simulate SPT survey, assuming Tinker et
al.~(2008) number density of clusters; the details are given in
section~\ref{res:counts}. The results for eRosita are expected to be
similar.

Typically, the cluster data alone is not enough to break the
degeneracies between different parameters and we need the constraints
from the primary CMB power spectrum. We add CMB data in the form of
scalar Cl's simulated as per Planck specifications.  We also apply a
Gaussian prior on $H_0$ as $H_0 = 70.4 \pm 3.6$ km/s/Mpc, where the
error is consistent with the recent results of the Hubble constant
from the SHOES (Supernovae and $H_0$ for the Equation of State)
program \citep{shoes}. We do not add any other possible future
datasets such as the supernovae or Baryon Acoustic Oscillations.

In analyzing the data, we keep $w_m \geq -0.1$ since we are interested
in constraining models which have sufficient amount of dark energy at
early times. The true $\Lambda$CDM model cannot be exactly reproduced
by the EDE class of models since $w_m \neq -1$, but in the limit $w_0
= -1, a_t \rightarrow 0 \ (z_t \rightarrow \infty), \Delta_t
\rightarrow 0$ these models replicate $\Lambda$CDM-like behavior.

As shown in figure~\ref{fig:SPT_mcmc}, CMB + cluster count analysis
can constrain the EDE parameters to $w_0 \lleq -0.9, a_t \lleq 0.2
(z_t \ggeq 4), \Delta_t \lleq 0.3$, which implies that cluster counts
+ CMB can constrain any deviation from $\ld$CDM behavior upto at least
z=4. For the SZ power spectrum data, we find that the EDE parameters
are constrained to $w_0 \lleq -0.85, a_t \lleq 0.15 (z_t \ggeq 5.5),
\Delta_t \lleq 0.15$ (see figure~\ref{fig:SZ_mcmc}). SZ power spectrum
provide complementary information about dark energy parameters
compared to cluster counts. For example, SZ power spectrum put tighter
constraints on the dark energy transition parameters compared to
cluster counts.  The matter density is reasonably reconstructed for
both the cluster counts and SZ power spectrum.

\section{Conclusions}\label{concl}

In this work, we have studied early dark energy models in light of
future galaxy cluster data. We study two EDE models, both of which are
allowed within the current observational constraints. The first model
(EDE1) transits from EDE behavior at redshift $z \ggeq 5$ to
$\ld$CDM-like behavior at present, while the second model (EDE2)
transits at a higher redshift ($z \simeq 9$) to an equation of state
$w_0 = -0.9$, which is close to but not identical to $\ld$CDM. For
both the models, as well as for the fiducial $\ld$CDM cosmology, we
show the predictions for cluster abundance and SZ power spectrum. For
that purpose, we consider two instruments -- eROSITA for X-ray
surveys, and ACT/SPT for SZ effect. We also do a likelihood analysis
of data simulated replicating the SPT and eRosita survey
specifications, along with the simulated Planck CMB data, to obtain
the constraints on the EDE parameters from future surveys, and find
that future galaxy cluster counts and SZ power spectrum can put
competitive constraints on these parameters.

It is worth noting some apparent differences between our work here,
and some other recent works on EDE signatures. For example, Fedeli et
al.~(2009) also considered several EDE models, but their cluster
counts are higher than for the fiducial $\Lambda$CDM model, although
$\sigma_8$ for EDE models are commonly lower in their work (as it is
the case here as well).  That is because for their choice of EDE
models Hubble parameters, $E(z)$ differs significantly from the
corresponding $\ld$CDM model. This affects minimum observable mass as
a function of redshift; their EDE models have much lower mass
threshold than $\Lambda$CDM, thus resulting in higher observable
cluster counts. In contrast, we have chosen models which have $E(z)$
similar to that of the corresponding $\ld$CDM model, to highlight the
difference coming from the perturbations rather than the background
expansion.  Similar difference from this work, due to the radical
departure of $E(z)$ from $\Lambda$CDM, can be seen in Waizmann \&
Bartelmann (2009), for the effects of EDE on SZ power spectrum. In
some other cases differences can arise depending whether one uses
low-k, CMB normalization of the power spectrum as done here, or
$\sigma_8$ normalization as in Sadeh et al.~(2007), although it is
clear that a successful model has to fulfill both constraints.

We show that the inclusion of dark energy perturbations has a major
effect on the matter power spectrum, therefore increasing the
possibility of discriminating EDE models from $\ld$CDM using large
scale structure probes. Neglecting the \de~\perts~leads to severe
underestimation of the imprint which the sharp transition in dark
energy equation of state leaves on the \de~\perts~and therefore on the
matter power spectrum.  We show that the models considered here which
are allowed by the current observations can be ruled out using the
future galaxy cluster probes. It is also interesting that both the
cluster counts and the SZ power spectrum are sensitive to the redshift
at which the transition between early and present day dark energy
occurs. We expect to put strong constraints on the equation of state
of dark energy at present using low redshift geometric observables
(such as the luminosity distance of type Ia SNe and Baryon Acoustic
Oscillations peaks). These probes of geometry of the Universe,
however, are insensitive to the high redshift behavior of dark
energy. The galaxy cluster observables such as their redshift
abundance and the SZ power spectrum, although at low redshift, are
sensitive to the perturbations in the early dark energy models, and
hence would be able to put constraints on the redshift of transition
from EDE behavior to present-day, $\ld$CDM-like behavior. With the
ongoing and future cluster surveys in microwave, optical and X-ray
waveband, galaxy clusters will be able to provide strong constraints
on the dynamical dark energy sector.

\section{Acknowledgements}

We thank Konstantin Borozdin, Salman Habib and Katrin Heitmann for
useful discussions.  We also thank the referee for his useful suggestions. The authors acknowledge support from Los Alamos
National Laboratory and the Department of Energy via the LDRD program
at LANL.

\end{document}